\documentclass{sf2a-conf}
\usepackage{graphicx}
\begin{document}

\TitreGlobal{SF2A 2008}

\title{Dust-enshrouded star formation in XMM-LSS galaxy clusters}
\author{Temporin, S.}\address{CEA Saclay, DSM/IRFU Service d'Astrophysique, Laboratoire AIM, CNRS, CEA/DSM, Universit\'e\ Paris Diderot, F-91191 Gif sur Yvette, France}
\author{Duc, P.-A.$^1$}
\author{Ilbert, O.}\address{Institute for Astronomy, Univerity of Hawaii, Honolulu, 96822 Hawaii}
\author{the XMM-LSS/SWIRE Collaboration}
\runningtitle{Dust-enshrouded star formation in XMM-LSS galaxy clusters}
\setcounter{page}{237}
\index{Temporin, S.}
\index{Duc, P.-A.}
\index{Ilbert, O.}

\maketitle
\begin{abstract} 

We present an investigation of the dust-enshrouded activity in a sample of X-ray 
selected clusters drawn from the XMM-LSS survey in the redshift range z $\sim$ 0.05 - 1.05. 
By taking advantage 
of the contiguous mid-IR coverage of the XMM-LSS field by the Spitzer SWIRE
legacy survey, we examined the distribution and number density of mid-IR bright
sources out to the cluster periphery and its dependence on redshift to probe 
the obscured side of the Butcher-Oemler effect. 
Toward intermediate redshift clusters we identified surprisingly high numbers
of bright 24 $\mu$m sources, whose photometric redshifts are compatible with 
cluster membership. The stacked surface density profile of 24 $\mu$m sources in clusters
within four redshift bins gives evidence for an excess of bright mid-IR sources 
at redshift z $\geq$ 0.4 at cluster-centric radii $\sim$ 200 - 500 kpc. 
Some traces of excess appear to be present at larger radii as well.
\end{abstract}
%
\section{Introduction}
  
Long after their collapse and the formation of the bulk of their stars, 
clusters of galaxies  still accrete new members. Subject to collisions and the
effects of the intracluster medium, the infalling galaxies loose their gas and 
ultimately their ability to form stars.
In the local Universe,it is now established that star formation is suppressed
in galaxy clusters. However, at z $\sim$ 1 a reversal trend has been recently observed,
with evidence of enhancement of star formation activity within higher density
environments (Elbaz et al. 2007; Marcillac et al. 2008).

At intermediate redshifts, observations with ISOCAM, onboard the ISO satellite,   
of a few individual galaxy clusters suggested the presence of an infrared
Butcher-Oemler effect (Fadda et al. 2000; Duc et al. 2002) and revealed the presence 
of several particularly 
active galaxies with total IR luminosities well above $10^{11}$ L$_{\odot}$ 
(Duc et al. 2004; Coia et al. 2005). 
If powered by 
dust-enshrouded star formation, as indicated by their spectra, these Luminous 
Infrared Galaxies (LIRGs) would exhibit star formation rates of several tens of solar 
masses per year - values that were unprecedented in a cluster environment.
The advent of the Spitzer Space Telescope with its wide field of view encouraged
further IR studies of galaxy clusters out to the cluster periphery (Geach et al. 2006;
Marcillac et al. 2007),
essentially confirming an important presence of luminous IR sources in clusters
at redshift z $\geq$ 0.4. A first attempt in quantifying an evolution with 
redshift has been done recently based on a collection of 8 massive clusters
in the range z = 0.02 - 0.8, observed at 24$\mu$m with Spitzer. This study indicates
an increasing fraction of mid-IR star-forming galaxies with redshift (Saintonge et al. 2008).

Here we present the first IR study based on an unbiased sample of X-ray selected clusters 
on a wide contiguous area of the sky covered with 24 $\mu$m Spitzer observations
to probe the obscured side of the Butcher-Oemler effect (Butcher \&\ Oemler, 1984)
at intermediate redshifts, out to the cluster periphery.

\section{The sample}

We considered X-ray selected clusters drawn from the first contiguous 5 deg$^2$ of 
the XMM Large Scale Structure (XMM-LSS)
survey (Pierre et al. 2004) and belonging to the XMM-LSS C1 sample.
These clusters obey precise selection criteria based on the properties of their
X-ray emission (Pacaud et al. 2007; Pierre et al. 2007) and constitute an unbiased sample
of groups and clusters (X-ray temperatures in the range 0.6 -- 4.8 keV) 
that spans the redshift range z $\sim$ 0.05 -- 1.05.
Part of our analysis included additional C1 clusters that were identified 
in the -- yet unpublished -- second  5 deg$^2$ of the XMM-LSS.
A wealth of ancillary multi-wavelength data is available for these clusters,
including either $u^{\ast}g^{\prime}r^{\prime}i^{\prime}z^{\prime}$ 
photometry\footnote{Based
on observations obtained with MegaPrime/Megacam, a joint project of CFHT and CEA/DAPNIA,
at the Canada-France-Hawaii Telescope (CFHT) which is operated by
the National Research Council (NRC) of Canada, the Institut National des Sciences de l'Univers
of the Centre National de la Recherche Scientifique (CNRS) of France, and the University of
Hawaii. This work is based in part on data products produced at TERAPIX and the Canadian 
Astronomy Data Centre as part of the Canada-France-Hawaii Telescope Legacy Survey, 
a collaborative project NRC and CNRS.} from the CFHT Legacy Survey (CFHTLS fields D1 and W1) 
or $g^{\prime}r^{\prime}z^{\prime}$ CFHT-Megacam photometry 
(from complementary observations of the northern part 
of the XMM-LSS field).
About 9 deg$^2$ of the XMM-LSS region of the sky were covered by observations 
with Spitzer Space Telescope in the four IRAC bands and in the three MIPS channels 
as a part of the SWIRE legacy survey (Lonsdale et al. 2003). This offered us the chance
to investigate the dust-enshrouded activity of galaxies out to large cluster-centric radii.
Spectroscopic observations necessary to cluster confirmation and
redshift determination were obtained by the XMM-LSS team during a series of dedicated
spectroscopic runs at various telescopes (see Table 2 of Pacaud et al. 2007 for details).
Furthermore, the XMM-LSS field includes the 02-hr field of the VIMOS-VLT Deep Survey (VVDS;
Le F\`evre et al. 2005).

From the C1 sample we selected a subsample of 32 clusters with 24$\mu$m and optical coverage.
For use as a control sample, we compiled a list of randomly selected fields in the XMM-LSS/SWIRE
area, by imposing as a constraint a separation of at least 6$^{\prime}$ from any catalogued X-ray 
source.  We selected all catalogued 24$\mu$m sources with flux F(24$\mu$m) $>$ 150 $\mu$Jy 
in the direction of these clusters and control fields within a radius of 10$^{\prime}$ from
either the X-ray position of a cluster or the central coordinates of a control field.
The only additional requirement was the presence of an optical counterpart to the 24$\mu$m 
sources, to reduce the contamination from spurious 24$\mu$m detections.
Total infrared luminosities of the sources were derived from their 24$\mu$m fluxes following
Chary \&\ Elbaz (2001) and assuming that the sources are all located at the cluster redshift.

\section{Surface density profiles of 24$\mu$m sources in individual clusters and control fields}

\begin{figure}[hb]
   \centering
   \includegraphics[width=\textwidth]{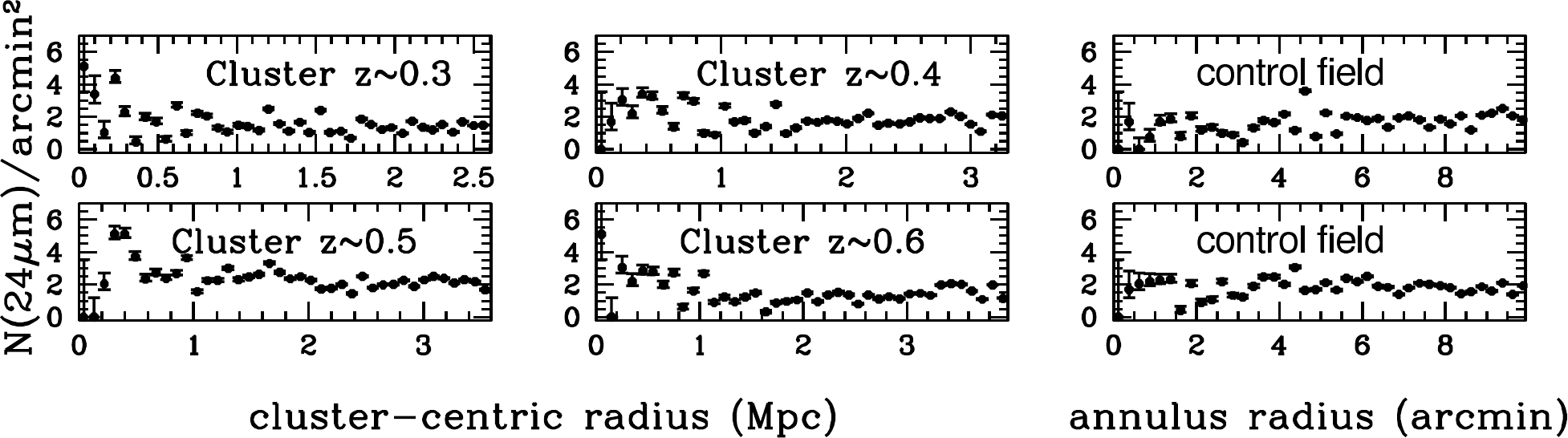}
      \caption{\emph{Left and middle column} -- Surface density profiles of 
      24$\mu$m sources computed in 
      0.25 arcmin-wide concentric annuli within 10$^{\prime}$ from the
      X-ray cluster position for a subsample of clusters at z $\sim$ 0.3 -- 0.6.
      \emph{Right column} -- The same quantity computed for two randomly selected
      control fields. Error bars are Poissonian.}
       \label{sd_individual}
   \end{figure}

As a first attempt in looking for any dependence of the distribution of mid-IR sources 
on the environment,
we examined the projected surface density of 24 $\mu$m sources as a function of
cluster-centric radius toward individual clusters in our sample.
Surface densities and associated Poissonian error bars were computed within 0.25 
arcmin-wide annuli centered on the X-ray position of the clusters.
We applied the same procedure to the control fields. 
Examples of profiles for a subsample of clusters and two control fields are shown 
in Fig.~\ref{sd_individual}.

This preliminary analysis showed some trends with radius of the projected 
surface densities of mid-IR sources and suggested the presence of an excess of sources in 
intermediate-redshift clusters at cluster-centric radii of a few hundred kpc to 1Mpc
with respect to the field.
However, the weakness of the signal and the high uncertainties due to 
the background fluctuations prompted us to build stacked density profiles
by averaging the signal that stems from different clusters.

\section{Stacked surface density profiles of 24$\mu$m sources}

\begin{figure}[ht]
   \centering
   \includegraphics[width=5.5cm]{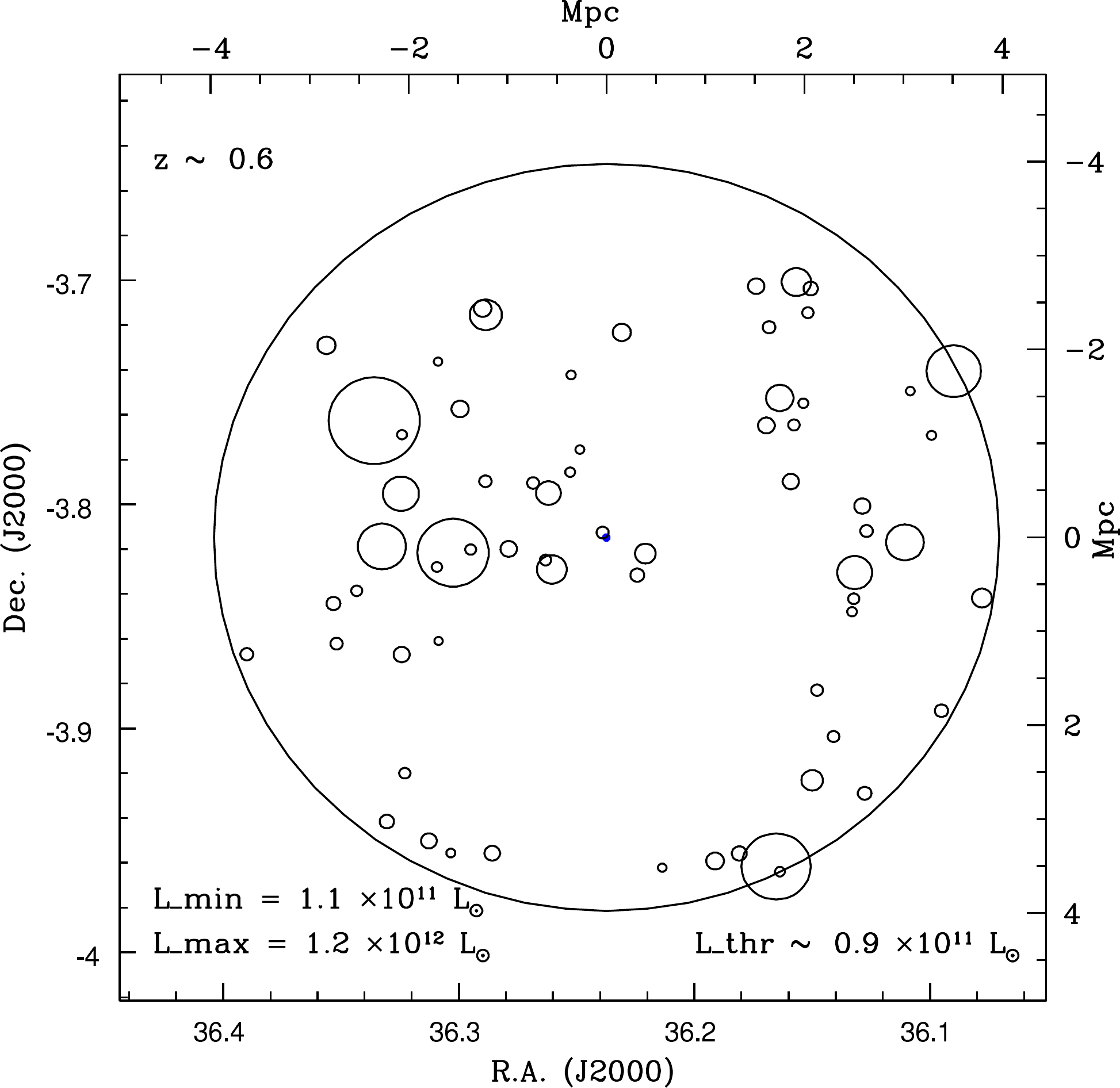}
   \includegraphics[width=5.5cm]{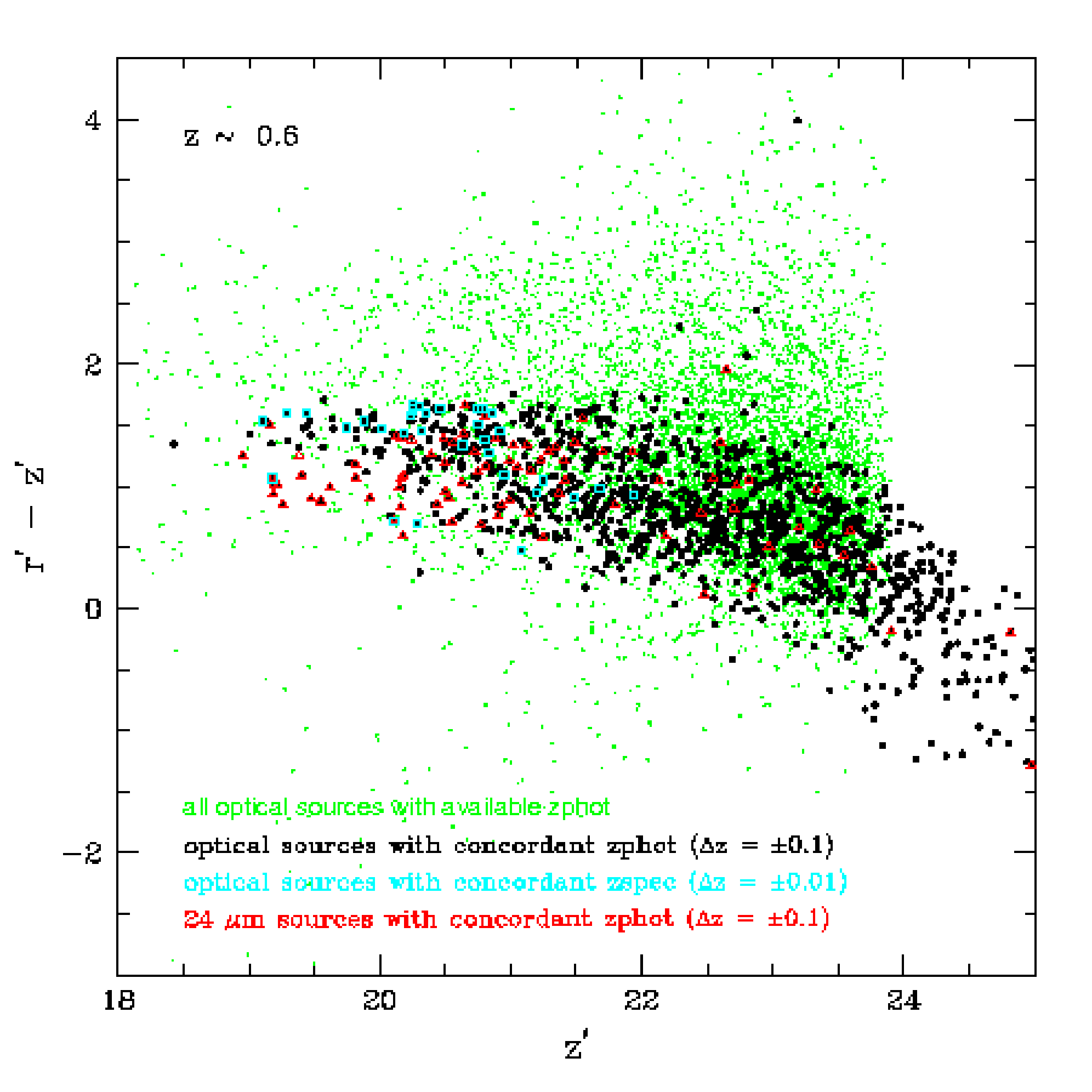}
   \includegraphics[width=5.5cm]{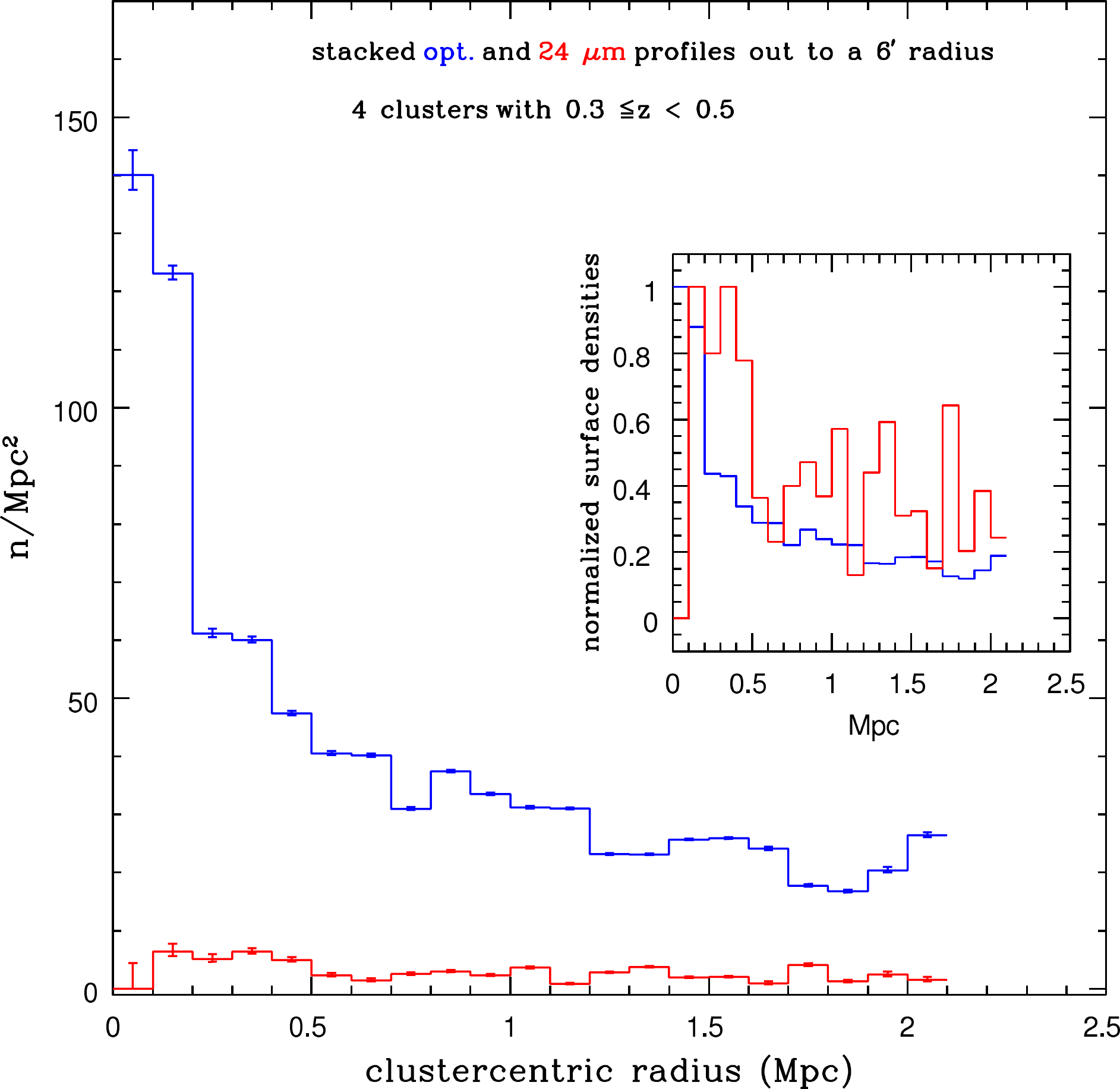}
      \caption{\emph{Left.} Example of the distribution of 24$\mu$m sources
      within a 10$^{\prime}$ radius for a cluster at z $\sim$ 0.6. The size of circles
      is proportional to the total IR luminosity of the sources.
      \emph{Centre.} Example of color-magnitude diagram for a cluster at z $\sim$ 0.6.
      Black dots and red triangles are concordant optical and 24$\mu$m sources,
      according to photometric redshifts. Green dots are all optical sources 
      in the direction of the cluster. Cyan squares are optical sources with 
      concordant spectroscopic redshift.
      \emph{Right.} Stacked surface density profile of concordant optical (blue) 
      and 24$\mu$m (red) sources -- according to photometric redshifts -- 
      for a subsample of 4 clusters in the redshift bin 0.3 -- 0.5. 
      The inset shows the same radial profiles normalized to the highest 
      peak in each distribution.}
       \label{opt-24mu}
   \end{figure}
   
In order to investigate the dependence on redshift of the observed 
density of mid-IR sources in clusters, we divided our sample into 
4 redshift bins, 0.05 $\leq$ z $<$ 0.3, 0.3 $\leq$ z $<$  0.5, 0.5 $\leq$ z $<$ 0.8, 
and 0.8 $\leq$ z $<$ 1.1.
For our stacking analysis we computed the surface density of 24$\mu$m sources 
within 0.1 Mpc-wide concentric annuli centred on the X-ray cluster position 
by combining all clusters within a given redshift bin. Thus, we obtained
a surface density radial profile of mid-IR sources for each redshift bin.
In the attempt to statistically account for the contamination by fore/background 
sources we built similar radial profiles by using our sample of control fields.
The shape of the radial profiles that included contaminant foreground/background sources
resulted to be strongly dependent on the redshift range and distribution of the 
considered fields (Temporin et al. 2008). 
Therefore, for correctly estimating the background, we randomly assigned 
redshifts to our control fields by reproducing the same redshift distribution
of our cluster sample. The comparison between the stacked radial profiles for
the clusters and the control fields showed the presence of a significant
excess of mid-IR sources at cluster-centric radii between $\sim$ 200 kpc and $\sim$ 1 Mpc
for clusters in the intermediate redshift bins.

This result was confirmed by the application of an alternative method to
take into account the contamination by fore/background sources.
This second method was based on a pre-selection of potential cluster 
members according to photometric redshifts. 
We used photometric redshifts obtained for the CFHTLS D1 field with the code Le\_Phare (Ilbert
et al. 2006) and complemented with additional photometric redshifts derived with the same 
method for a portion of the W1 field.
Sources were considered as candidate cluster members when their photometric
redshift was compatible within errors with the available estimate of cluster redshift.
We additionally considered a $\Delta$z = 0.01 to take into account the velocity 
dispersion of galaxies within a cluster (including the cluster periphery to which our
analysis extends).

We found an unexpectedly high number ($\sim$ 20 to 100) of 24$\mu$m sources having 
photometric redshifts compatible with cluster membership, especially for clusters 
in the intermediate/higher redshift bins, out to 10$^{\prime}$ radii from the cluster centre.
Out of these mid-IR sources, the number of luminous infrared galaxies (LIRGs) ranges
between a few for clusters at z $\leq$ 0.3 and $\sim$ 60 -- 100 for clusters at
z $\geq$ 0.5. An example plot showing the distribution of mid-IR sources
across an intermediate-redshift cluster is shown in Fig.~\ref{opt-24mu} (left hand). We note
that the brightest sources tend to avoid the very centre of the cluster.

Also, in Fig.~\ref{opt-24mu} (centre) we show an example of color-magnitude diagram 
for one of the clusters in our sample. The effectiveness of the photometric
redshift pre-selection
in reducing the fore/background contamination is evident. Interestingly,
as expected for dusty star-forming galaxies,
the mid-IR sources are mainly distributed within the ``blue cloud'' of galaxies and in the
region of the diagram between the ``blue cloud'' and the red sequence, the so-called
``green valley'', while only a small number of them is found in the red sequence.

As expected, the stacked radial profiles of the control fields -- after the photometric
redshift pre-selection of sources -- resulted in a flat distribution of the surface
density of 24 $\mu$m sources. 
Conversely, the stacked cluster radial profiles showed significant peaks in the 
distribution, again hinting at an excess of sources at radii 200 -- 500 kpc and $\sim$ 1 Mpc for
the intermediate redshift bins. Lower significance peaks of density are seen also at 
larger radii, toward the cluster periphery.
A comparison with the radial distribution of (pre-selected in photometric redshift) 
optical sources allowed us to exclude that the observed 24$\mu$m sources density peaks
were just mirroring the general galaxy distribution in clusters.
A comparison between the stacked radial distributions of optical and 24$\mu$m sources is
shown in Fig.~\ref{opt-24mu} (right hand) for a subsample of 4 clusters in the redshift bin
z = 0.3 -- 0.5.

\section{Conclusions}

Our statistical analysis of the distribution of 24$\mu$m sources in an unbiased sample of
X-ray selected clusters drawn from the XMM-LSS survey has revealed the presence
of a significant excess of mid-IR sources that are compatible with cluster membership
at cluster-centric radii of 200 -- 500 kpc and $\sim$ 1 Mpc in the intermediate
redshift bins of our sample (z $>$ 0.3). At these redshifts, our analysis extends
out to $\sim$ 3 to 4 Mpc radii from the cluster centre, as defined by the X-ray emission,
and, with the adopted flux threshold F(24$\mu$m) $>$ 150 $\mu$Jy,
the identified sources fall mostly in the LIRG regime. 
The brightest mid-IR sources tend to avoid the cluster centre. 
Our results are understood as a signature of the IR Butcher-Oemler effect in clusters.
The detailed analysis of the whole sample is presented elsewhere (Temporin et al. 2008).

\section*{Acknowledgements}
We are grateful to T. Evans and M. Polletta for providing us with the latest 
version of the band-merged SWIRE catalogue for the XMM-LSS field.
This work is based on observations
made with the Spitzer Space Telescope, which is operated by the Jet Propulsion
Laboratory, California Institute of Technology under NASA contract 1407.


%





\end{document}